\documentclass[twoside]{dis07}
\usepackage[latin1]{inputenc}
\usepackage[dvips]{graphicx,epsfig,color}
\usepackage{wrapfig,rotating}
\usepackage{amssymb,amsmath,array}

\begin{document}

\title{Global QCD Analysis and Collider Phenomenology---CTEQ }
\author{Wu-Ki Tung$^{1,2}$, H.L.~Lai$^{1,2,3}$, J.~Pumplin$^1$, P.~Nadolsky$%
^4$, and C.-P.~Yuan$^1$. \vspace{.3cm} \\
$^1$ Michigan State University, East Lansing, MI - USA \vspace{.1cm}\\
$^2$ University of Washington, Seattle, Washington - USA \vspace{.1cm}\\
$^3$ Taipei Municipal University of Education, Taipei, Taiwan \vspace{.1cm}\\
$^4$ Argonne National Laboratory, Argonne, IL, USA\\
}
\maketitle

\begin{abstract}
An overview is given of recent progress on a variety of fronts in the global
QCD analysis of the parton structure of the nucleon and its implication for
collider phenomenology, carried out by various subgroups of the CTEQ
collaboration.

\end{abstract}



\section{Introduction}

Parton distribution functions (PDFs) are essential input to all calculations
on high energy cross sections with initial state hadrons. \ PDFs are
extracted from comprehensive \emph{global analysis} of available hard
scattering data within the framework of \emph{perturbative QCD.} This report
covers recent progress on global QCD analysis made by members of the CTEQ
collaboration on a variety of fronts.~\cite{url}

The basis of most of the recent progress is a new implementation of the
general mass (GM) formulation for perturbative QCD that systematically
includes heavy quark mass effects, both in kinematics and in the
order-by-order factorization formula.~\cite{Tung:2006tb} The next section
describes the main implications of the new global QCD analysis on collider
phenomenology at the Tevatron and the LHC.~\cite{YuanEw}

This is followed by the first in-depth study of the strangeness sector of
the parton parameter space, based on the most up-to-date global analysis.~%
\cite{Lai:2007dq} We found that current data imply a symmetric component of the
strange parton distribution, $s(x)+\bar{s}(x)$, that has a shape independent of
that of the
isospin singlet non-strange sea; and a strangeness asymmetry function $s(x)-%
\bar{s}(x)$ that has a slightly positive first moment.

The same formalism has been applied to investigate the possibility of a
non-perturbative (intrinsic) charm component in the nucleon.~\cite%
{Pumplin:2007wg} This study is discussed in a separate talk in this
workshop \cite{TungHQ}. In a significant expansion of global QCD analysis,
we have succeeded in combining the traditional fixed-order global fits with $%
p_{t}$ resummation calculations. \ Combined conventional and $p_{t}$%
-resummed global fits can now be made to pin down parton degrees of freedom
that are most pertinent for precision $W$-mass measurement and Higgs particle
phenomenology.~\cite{YuanPt} Another subgroup of CTEQ has performed a
detailed investigation of the role of recent neutrino scattering experiments
(NuTeV, Chorus) and fixed-target Drell-Yan cross section measurement (E866)
on global analysis, particularly pertaining to the large-$x$ behavior of
parton distributions. The results are reported in \cite{Owens:2007kp}.

Due to space limitation, it is impossible to include in this short written
report the figures that illustrate the results discussed in the corresponding
talk, as summarized above. However, since the slides for the talk have been
made available at the official conference URL \cite{url}, we shall make use of
these, and refer the reader to the actual figures by the slide numbers where
they appear in the posted talk \cite{url}. The same space limitation restricts
citations to only the papers and talks on which this report is directly based.

\section{New Generation of PDFs and Their Implications for Collider
Phenomenology}

The base parton distribution set from the new generation of global analysis
incorporating the GM formalism for heavy quark mass effects is the CTEQ6.5M
PDF set \cite{Tung:2006tb}. The main improvements over the previous
generation of PDFs---CTEQ6.0 and CTEQ6.1---are the mass treatment and the
incorporation of the full HERA Run 1 cross section measurements, with their
correlated systematic errors.

The most noticeable change in the output parton distributions is a sizable
increase in the $u$- and $d$-quark distributions in the region $x\sim%
10^{-3}$ for a wide range of $Q$. The three figures on slides 4/5 of \cite%
{url} show the ratio of the CTEQ6.1 $u,d$-quark and the gluon distribution to
that of CTEQ6.5 at $Q=2$ GeV. The differences are moderated by QCD evolution,
but still persist to a high energy scale such as the W/Z masses. The origin of
these differences can be traced to the treatment of the heavy quark mass, as
explained in slide 6. This change has immediate phenomenological consequences.
\ The figure on slide 7 shows ratios of predicted cross sections at the LHC for
the standard model (SM) processes $W^{\pm }/Z$ production, Higgs production
$gg\rightarrow H^{0}$, and associated production of $W^{\pm }H$; as well as
some representative beyond standard model (BSM) processes, e.g.\ charged Higgs
production $\bar{s}c\rightarrow H^{+}\rightarrow \bar{b}t$.

Of immediate interest is the 7\% increase in the predicted W and Z production
cross sections at LHC (which are sensitive to PDFs in the $x\sim%
10^{-3} $ range) compared to previous estimates. The plot on slide 10 shows
the predicted Z vs.~the W cross sections for several commonly available PDF
sets. The predictions seem to fall into two groups, with no obvious pattern.
The results on slide 11 represent an attempt to see whether the difference
between Zeus and H1 predictions can be reproduced in the CTEQ framework. We
do not see a substantial difference between the two experimental inputs, but
do see a clear dependence on how mass effects are treated. Further mysteries
are (i) why are the Zeus predictions independent of their mass treatment;
and (ii) why are their predictions with mass effects so different from that
of MRST, even though they use the MRST formalism for mass treatment. The resolution of
these apparent puzzles concerning the W and Z cross sections at the LHC is
clearly of great importance to its physics program.

To see the impact of the new PDFs on collider phenomenology in general, it is
convenient to examine the luminosity curves. These are shown in slides 8-9 over
the range 10 GeV $<\hat{s}<$ 5 TeV for LHC (normalized to that of CTEQ6.1),
including bands representing the estimated uncertainties due to experimental
input to the global analysis. The quark-quark ($q\text{-}q$) luminosity curves
show the largest change between the two generations of PDFs; the $g\text{-}g$
luminosity is shifted only slightly, and the $g\text{-}q$ luminosity shift lies
in between.

The cross sections shown in slide 7 also include some typical BSM processes.
Notice in particular the very large predicted cross section for the last
process due to a new PDF set CTEQ6.5C that permits a non-perturbative
(intrinsic) charm component of the nucleon \cite{Pumplin:2007wg}.

In the base PDF set CTEQ6.5M, we adopted the conventional assumptions that the strange
distributions $s(x)$ and $\bar{s}(x)$ are of the same shape as the isospin symmetric
non-strange sea at the initial scale $\mu =Q_{0}$ for QCD evolution, and that the charm distribution
$c(x)$ is zero at the scale $\mu =m_{c}$. There are of no independent degrees of freedom for
strange and charm.
\ The improved
theoretical and experimental inputs to the new generation of global analysis
now permit us to relax these ad hoc assumptions, and hence to study where the
truth lies. The results on strange PDFs obtained by \cite{Lai:2007dq} will be
summarized in the following section. The subject of charm PDF is covered in \cite%
{TungHQ}.
\section{Systematic Study of the Strangeness PDFs}

Within the global QCD analysis framework, the only currently measurable
process that is directly sensitive to the strange distributions $s(x)$ and $%
\bar{s}(x)$ is dimuon (charm) production in neutrino (and anti-neutrino)
scattering off nucleons, via the partonic process $W^{+}(W^{-})+s\,(\bar{s}%
)\rightarrow c\,(\bar{c})$. The final data of the NuTev experiment \cite%
{Mason:2006qa} is thus crucial for this analysis. The constraining power of
these data can be realized, however, \emph{only within the framework of a
comprehensive global analysis}, since the same final state is produced also
by the down quarks, and since the strange sea is intricately coupled to the
gluon and the non-strange partons by QCD interactions. Also, because of the
presence of the charm particle in the final state of the dimuon signal, a
consistent theoretical treatment of heavy quark mass effects for both
charged-current and neutral-current DIS processes \cite{Tung:2006tb} is
essential to obtain reliable results.

For convenience, we define the symmetric strange sea\textbf{\ }$%
s_{+}(x)\equiv s(x)+\bar{s}(x)$ and the strangeness asymmetry $%
s_{-}(x)\equiv s(x)-\bar{s}(x)$. All these functions refer to the initial
distributions at $\mu =Q_{0}$; QCD evolution then dictates their $\mu $
dependence at higher energy scales. We address the following three issues in
turn.

\textbf{Is the shape of the symmetric strange sea independent of that of the
non-strange sea?} The answer appears to be yes, according to the up-to-date
global analysis \cite{Lai:2007dq}. The evidence is shown on slide 14. The
table gives the changes in the goodness-of-fit for the full set of 3542 data
points used in the global analysis, as well as the subset of 149 points for
neutrino dimuon data sets, that we found in performing a series of global
analysis, using 2/3/4 independent strangeness shape parameters, compared to
the CTEQ6.5M reference fit that tied the shape of $s(x)$ and $\bar{s}(x)$ to
that of the non-strange sea. We see that there is a substantial improvement
in the quality of the fit to the dimuon data with $s_{+}(x)$ different from
that of the non-strange sea. We also see that current data cannot
discriminate between 2-, 3-, or 4-parameter forms for $s_{+}(x)$. Thus, a
2-parameter form will serve as a practical working hypothesis.

\textbf{What is the size of the symmetric strange sea, and what are the
allowed ranges for its size and shape}? Slide 15 presents results of our
study on these issues. The upper figure shows the goodness-of-fit in terms
of $\chi ^{2}/$point for the dimuon data (deep parabola) and for the global
data (shallow parabola) as a function of the momentum fraction carried by
the strange sea, $\langle x\rangle _{s+}=\int xs_{+}(x)dx$. The dimuon data
clearly favor a central value of $\langle x\rangle _{s}\sim 0.027$. The
range of allowed size is obtained by adopting a 90\% confidence level
criterion. In terms of the ratio of the first moments (fractional momentum)
of the strange to non-strange sea, this range corresponds to ($0.27,$~$0.67$%
), as indicated on the slide. \ The range of possible shape of $s_{+}(x)$ is
a little more elusive to quantify. The lower figure presents a range of
possible candidates, within the 90\% C.L.~criterion, when both the size and
shape parameters are allowed to vary. These representative PDF sets are
labeled CTEQ6.5S$n$, $n=0,1,...,4$, with $n=0$ being the central fit.

\textbf{Current status of the strangeness asymmetry:} Non-perturbative
models of nucleon structure suggest a possible non-vanishing strangeness
asymmetry. Within the PQCD framework, QCD evolution beyond the first two
leading orders causes a non-vanishing $s_{-}(x,\mu ),$ even if one starts
with a symmetric strange sea. Historically, $s_{-}(x)$ was first studied
phenomenologically in 2003 as a possible explanation for the
\textquotedblleft NuTeV anomaly\textquotedblright\ associated with the
Weinberg angle measurement. Therefore, it is natural to ask: what can we say
about $s_{-}(x)$ currently, now that both the theory and experimental situation
have improved? \ The results of our study, \cite{Lai:2007dq}, are summarized
in slide 17: (i) current global analysis still does not require a non-zero $%
s_{-}(x)$, although it is consistent with one; (ii) the best fit corresponds
to a positive asymmetry $\langle x\rangle _{s-}=\int xs_{-}(x)dx\sim 0.002$;
and (iii) the 90\% C.L. range for $\langle x\rangle _{s-}$ is ($-0.001,~0.005
$). These results are consistent with both the 2003 CTEQ study and the most
recent NuTeV analysis \cite{Mason:2006qa}. The figures on slide 17 show the
shape of $s_{-}(x)$ and the momentum distribution $xs_{-}(x)$ for a variety
of possible candidate PDFs within the 90\% C.L. criterion.

\section{New neutrino DIS and Drell-Yan data and large-x PDFs}

It has been known for some time that the relatively recent NuTeV total cross
section and E866 Drell-Yan cross section data sets pose puzzling dilemmas
for quantitative global QCD analysis of PDFs, as indicated in slides 20 and
21. Attempts to incorporate these data in global analysis by Owens \textit{%
et.al.}~\cite{Owens:2007kp} led to the following key observations: (i) the NuTeV data set pulls against
several of the other data sets, notably the E-866 and the BCDMS and NMC data.
Nuclear corrections (heavy target) do not improve the situation. (In fact,
assuming no nuclear correction lessens, but does not remove, the problem.); (ii) the
conflicts are most pronounced when one examines the $d/u$ ratio. Adding
NuTeV and E-866 simultaneously in the global analysis causes the $d/u$ ratio
to flatten out substantially, resulting in worsened fits to other precision
DIS data; and (iii) the E866 $pp$ data is more comparable with precision DIS
data sets than the $pd$ data. Slides 23 - 26 show the figures that support
these observations.

\medskip
\noindent {\large \bf Conclusion:} Results presented here, in conjunction with
those covered in \cite{YuanEw,TungHQ,YuanPt}, represent significant
evolutionary advancement of global QCD analysis, as well as some
ground-breaking development (such as the incorporation of $p_t$ resummation
\cite{YuanPt}). There are, however, also open problems that require further
study and resolution \cite{Owens:2007kp}. Much remains to be done.



\end{document}